\documentclass{article} 
\usepackage[most]{tcolorbox}
\usepackage{enumitem}
\usepackage{amsmath} 

\usepackage{authblk}

\title{Zero-Cost Detection of Fabricated Citations: A Staged Heuristic Approach for Computing Dissertation Assessment} %

\author[1]{Hazrat Ali}
\author[2]{Mohamed Elawady}
\affil[1]{University of Stirling, Stirling, United Kingdom}
\affil[2]{University of Strathclyde, Glasgow, United Kingdom}

\date{\today} 

\definecolor{greencolor}{RGB}{39, 174, 96}       
\definecolor{bluecolor}{RGB}{52, 152, 219}

\begin{document} 
\maketitle 

\begin{abstract}
Generative AI's growing role in academic writing has driven a rise in fabricated references across dissertations and student reports. We present the Fabrication Severity Stack (FSS), a three-tiered framework classifying citation fabrications by detection complexity - from non-existent references to subtle content misattribution.
Building on the FSS, we propose a staged heuristic \textbf{verification workflow}, executable through mainstream LLM interfaces. We argue that this framework offers computer science educators/examiners a practical, zero-cost basis for embedding reference verification into the assessment of dissertation and students' reports. In a pilot study on 100 citations (20\% seeded with Level 1 and Level 2 fabrications), the proposed heuristic workflow enabled mainstream LLM interfaces to correctly flag 9–10 of 10 non-existent references and partially detect metadata corruption. These preliminary results suggest that freely accessible, general-purpose LLMs can achieve high recall on citation verification without reliance on proprietary detection tools.
\end{abstract}

\section{Fabrication Severity Stack}
Citation of relevant literature lies at the core of academic scholarship and students' writing, serving as the foundation for validating claims and anchoring discoveries. However, as Generative AI increasingly becomes a ubiquitous tool in academic writing, a critical vulnerability has emerged: the proliferation of ``fabricated citations'' within dissertations, student reports, and beyond. Dissertations and reports are increasingly being authored with LLM assistance, specifically during literature review and related-work drafting -- precisely the stages most susceptible to fabricated citations. Hence, computer science educators/examiners who evaluate these documents therefore need a systematic vocabulary and a repeatable verification procedure, rather than relying on spot-checks or intuition.

Independent audits corroborate the scale of the problem. For example, a systematic comparison of ChatGPT-generated bibliographies found that a substantial share of references were either entirely non-existent or contained material citation errors such as incorrect authorship, wrong publication years, or fabricated DOIs \cite{walters2023fabrication}. Even peer-reviewed venues are not immune. An analysis by Topaz et al. found more than 4,000 fabricated references in studies published betweeen Jan 2023 and Feb 2026 \cite{topaz2026fabricated}. An audit of ACL Anthology papers documented a rise in hallucinated references between 2024 and 2025 \cite{sakai2026hallucitation}. Of 700,000 citations in submissions made to ICLR 2026\footnote{International Conference on Learning Representations}, GPTZero flagged 9,000 citations that required additional manual checking \cite{naddaf2026growing}. 

In this context, this work categorizes AI-generated reference fabrications into a three-layered hierarchy based on the complexity and severity of their detection. We conceptualize this taxonomy as the \textbf{Fabrication Severity Stack} (FSS), a structured framework for analysing and understanding fabricated citations in computer science dissertations, students' reports, and beyond. 
The \textbf{FSS}-based categorization consists of the following three levels:\\
\textbf{Level 1 -- Syntactic Non-Existence.} Total fabrication or absolute LLM hallucinations of titles, DOIs, or papers that do not exist at all. These are often easily identifiable via automated metadata lookups, DOI verification (e.g., on Crossref), and direct web searches.\\
\textbf{Level 2 -- Metadata Corruption.} Partial hallucinations where valid metadata elements are blended with fabricated ones. Their detection might pose a medium-complexity challenge and require cross-referencing multiple metadata fields to flag inconsistencies.\\
\textbf{Level 3 -- Content Corruption.} The paper exists, and all metadata is perfectly valid, but the cited text does not support the author's claim (e.g., fabricated / selective quotation, or misattributed findings). Detection of such cases might be highly complex and often requires full-text retrieval, semantic analysis, or advanced RAG-based tools.

\begin{figure}[t!]
\centering
\includegraphics[width=0.72\linewidth]{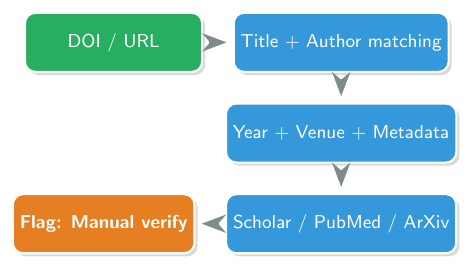}
\caption{A heuristics-based verification workflow.}
\label{fig:workflow}
\vspace{-5pt}
\end{figure}

\section{Heuristic-Based Workflow} We outline a lightweight, three-stage pipeline that computer science educators can execute manually or semi-automate with a general-purpose LLM interface, without requiring a bespoke detection tool (also see Fig.~\ref{fig:workflow}):

\begin{itemize}[leftmargin=*]
\item \textit{Existence Check.} For each reference, resolve the DOI or URL against an authoritative indexer (Crossref, DBLP, arXiv, PubMed, or Google Scholar). A reference is flagged immediately if the DOI does not resolve, the URL returns a dead link, or no matching record can be located under any permutation of the stated title. This single check disposes of the majority of total fabrications.

\begin{tcolorbox}[colframe=greencolor, colbacktitle=greencolor, title=Sample Prompt]
For every citation below, determine whether it corresponds to a real, locatable publication. Attempt to resolve the DOI/URL if present, and
search Crossref, DBLP, arXiv, PubMed, or Google Scholar for the title under reasonable variations (typos, subtitle differences, translated
titles). Output one of: EXISTS (with source + link), NOT FOUND. If EXISTS, note whether
the found record matches the title exactly or only approximately.

\end{tcolorbox}

\item \textit{Metadata Consistency Check.} For references that pass the existence check, cross-validate author list, publication year, venue, and volume/page numbers against the indexer record. The aim is to identify specific metadata discrepancies, for example; a real paper title attached to the wrong author list, or a plausible-sounding venue that never published the cited study. Automating these checks thus reveals metadata blending and warrants closer inspection.

\begin{tcolorbox}[colframe=bluecolor, colbacktitle=bluecolor, title=Sample Prompt]
For every citation below, check the fields (authors, year, venue, volume, pages) and output MATCH, MISMATCH, or PARTIAL per field, plus
a one-line reason for any non-match. Do not judge the paper's content.
Output as a single table with columns:\\
Reference | Field | Status | Reason

\end{tcolorbox}

\item \textit{Claim-Support Check.} For references surviving the first two stages, sample a subset of citations (prioritizing those supporting the dissertation's central claims) and retrieve the actual cited passage or abstract. An LLM, or the examiner, then assesses whether the passage substantiates the specific claim attributed to it, distinguishing genuine support from selective quotation, overgeneralization, or outright misattribution. Because this stage requires semantic judgment rather than lookup, it is relatively labor-intensive and should be applied selectively rather than exhaustively.

\end{itemize}

\textbf{Preliminary results:}
To assess the practical viability of the proposed heuristic workflow, we conducted an initial pilot study on a curated set of 100 citations drawn from student dissertation bibliographies, of which 20 (20\%) were deliberately seeded with fabrications spanning Level 1 (non-existent references) and Level 2 (metadata-corrupted references) of the FSS, in equal proportion. Two mainstream LLM interfaces, \textit{Claude, Sonnet 5} and \textit{Gemini, Flash 3.5}, were evaluated using the existence-check and metadata-consistency prompts described above. Claude correctly flagged all 10 Level 1 fabrications, and additionally surfaced several Level 2 cases by identifying mismatched author lists and blended metadata fields, indicating that a single structured pass can partially generalize beyond its targeted severity tier. Gemini correctly flagged 9 of the 10 Level 1 fabrications, missing one case where the fabricated title closely resembled a real publication. While this pilot is small in scale and does not yet constitute a large scale benchmark, it suggests that mainstream, freely accessible LLM interfaces — without bespoke detection infrastructure — can achieve high recall on citation verification, lending preliminary support to the idea that expensive proprietary tools may not be a strict prerequisite for effective citation screening.

\section{Final Remarks}
Rigorous citation verification has become an indispensable component in assessments. The rapid evolution of LLM technology has led to growth in specialized detection tools.\footnote{Examples include: GPTZero, SwanRef, CiteTrue, CiteMe, Citely, CiteCheck, CiteSure, and MisciteBench. More updates on https://github.com/anonymized/anonymized.}. Academics might not need to rely on expensive detection tools, as comparable outcomes can be achieved by employing structured heuristic workflows via mainstream LLM interfaces. This approach centers on automating the verification of citations for: (i) DOI/URL validity, (ii) title/author consistency, and (iii) metadata alignment across indexers (PubMed/arXiv/Scholar). References failing any of these automated or semi-automated checks should be flagged for manual review rather than automatically rejected, since indexer coverage gaps and formatting idiosyncrasies can produce false positives. Applying stages (i) and (ii) across an entire bibliography is inexpensive and should be treated as a crucial pre-submission step for dissertations and theses, analogous to running a plagiarism check. Stage (iii), by contrast, is best reserved for spot-checking claims, since Level 3 fabrications are structurally distinct from Levels 1 and 2 (the reference itself is entirely legitimate) and consequently present a substantially greater diagnostic challenge that current automated tools handle unreliably \cite{sakai2026hallucitation}. We acknowledge the preliminary nature and limited scale of the pilot study.

\vspace{-5pt}

\bibliographystyle{plain}
\bibliography{refs}

\end{document}